\documentclass[number=noenddot]{spie}  %>>> use for US letter paper

\usepackage[]{graphicx}
\usepackage{amsmath}

\usepackage{caption}
\usepackage{color}
\title{Development of a New, Precise Near-infrared Doppler Wavelength Reference: A Fiber Fabry-Perot Interferometer} 

\author{Samuel Halverson\supit{a},  Suvrath Mahadevan\supit{a,b}, Lawrence Ramsey\supit{a,b}, Stephen Redman\supit{c}, Gillian Nave\supit{c} , John C. Wilson\supit{d}, Fred Hearty\supit{d}, Jon Holtzman\supit{e}
\skiplinehalf
\supit{a}Department of Astronomy \& Astrophysics, The Pennsylvania State University, 525 Davey Laboratory, University Park, 16802, USA; \\
\supit{b}Center for Exoplanets \& Habitable Worlds, The Pennsylvania State University, University Park, PA 16802; \\
\supit{c}Atomic Physics Division, National Institute of Standards and Technology, Gaithersburg, MD 20899, USA; \\
\supit{d}Department of Astronomy, University of Virginia, P.O. Box 400325, Charlottesville, VA 22904-4325, USA; \\
\supit{e}Department of Astronomy, New Mexico State University, Box 30001, Las Cruces, NM 88003, USA; \\
}

\authorinfo{E-mail: shalverson@psu.edu, Telephone: 1 650 392 5909}
 
\begin{document} 
  \maketitle 

\begin{abstract}
We present the ongoing development of a commercially available Micron Optics\footnote[1]{Certain commercial equipment, instruments, or materials are identified in this paper in order to specify the experimental procedure adequately. Such identification is not intended to imply recommendation or endorsement by the National Institute of Standards and Technology, nor is it intended to imply that the materials or equipment identified are necessarily the best available for the purpose.} fiber-Fabry Perot Interferometer as a precise, stable, easy to use, and economic spectrograph reference with the goal of achieving $<$1 m/s long term stability. Fiber Fabry-Perot interferometers (FFP) create interference patterns by combining light traversing different delay paths. The interference creates a rich spectrum of narrow emission lines, ideal for use as a precise Doppler reference. This fully photonic reference could easily be installed in existing NIR spectrographs, turning high resolution fiber-fed spectrographs into precise Doppler velocimeters. First light results on the Sloan Digital Sky Survey III (SDSS-III)  Apache Point Observatory Galactic Evolution Experiment (APOGEE) spectrograph and several tests of major support instruments are also presented. These instruments include a SuperK Photonics\footnotemark[1] fiber supercontinuum laser source and precise temperature controller. A high resolution spectrum obtained using the NIST 2-m Fourier transform spectrometer (FTS) is also presented. We find our current temperature control precision of the FFP to be 0.15 mK, corresponding to a theoretical velocity stability of 35 cm/s due to temperature variations of the interferometer cavity.

\end{abstract}

\keywords{wavelength references, high resolution spectroscopy, near-infrared spectrograph design, extra-solar planets, radial velocity measurements}

\section{INTRODUCTION}
\label{sec:intro}  % \label{} allows reference to this section

Modern day Doppler radial velocity (RV) surveys require precise spectrographs to measure the minute spectral shifts in stellar spectra induced by unseen companions. To date, over 500 planets have been discovered using the RV method. Current dedicated planet hunting spectrographs can achieve $\sim$1 m s$^{-1}$ precision on bright sun-like stars with high cadence\cite{2011A&A...534A..58P}. This high precision allows for the detection of a wide range of low mass planets. The detection of these planets requires stable, accurate wavelength calibration devices for confident detections over long periods. 

A new area of exploration lies in the selection of stars studied in RV surveys. The overwhelming majority of planets discovered have been found orbiting stars of similar spectral type to our own Sun (FGK). These stars are bright in the visible and generally well understood, but do not comprise the majority of stars in our Galaxy. M-dwarfs are estimated to makeup the majority (over 70 percent) of stars in the Milky Way. These small stars have low masses, resulting in larger Doppler reflex signals when gravitationally interacting with an unseen planetary mass companion. M-dwarf Habitable Zones (HZ) are also much closer to the host stars in M-dwarf systems than larger solar type stars, making detection of low-mass habitable planets more feasible. M-stars are brightest in the NIR bands however, where few precise and dense wavelength references for spectrograph calibration exist.

Iodine (I$_2$) cells have been used to obtain precise velocity measurements on F,G, and K stars \cite{2011ApJ...730...10H} in the 400 - 700nm wavelength regime, but lack any significant spectral features in the NIR to precisely measure low amplitude ($<$5  m s$^{-1}$) velocities.
Similarly, the popular Th/Ar emission lamp has been utilized to achieve sub m s$^{-1}$ long term precision on Solar type stars using stabilized spectrographs\cite{2011arXiv1109.2497M}, but does not contain enough lines beyond 1000 nm to be an optimal NIR reference source.

With the slew of current and next generation high-resolution NIR spectrographs (e.g. HPF: Habitable-zone Planet Finder\cite{2010SPIE.7735E.227M}, CARMENES: Calar Alto high-Resolution search for M dwarfs with Exoearths with Near-infrared and optical Echelle Spectrographs, \cite{2011IAUS..276..545Q} APOGEE: Apache Point Observatory Galactic Evolution Experiment\cite{2010SPIE.7735E..46W}, iShell\cite{2008SPIE.7014E.208T}, CRIRES: Cryogenic high-resolution Infrared echelle Spectrograph\cite{2004SPIE.5492.1218K}), development of spectrograph references rich with spectral features in this wavelength regime is therefore critical for the advancement of NIR Doppler velocimetry.

Fabry-Perot interferometers (FPI) create emission spectra by interfering light traversing separate delay paths. The resulting interference spectrum provides a rich distribution of narrow lines, ideal for use as a spectrograph reference. FPIs have been tested
as wavelength references in the past with excellent short term stability\cite{2010SPIE.7735E.164W}, but have yet to show the required long term stability for precision measurements down to the m s$^{-1}$ level.

We present preliminary tests of a commercially available H-band (1500 - 1700 nm) Fiber Fabry-Perot
(FFP) designed by Micron Optics\footnotemark{}. The FFP is a single-mode fiber (SMF) assembly
with a pair of reflective dielectric stack mirrors spliced into the beam-path. This simple design, when compared
to the standard FPI cavity, yields a more stable device over the long term. If shown to be a stable
and precise reference, the FFP design could be used on any fiber-fed spectrograph, either
in the optical or the NIR.

In Section~\ref{sec:wvl_ref} we summarize existing NIR and optical wavelength references for comparison purposes. In Section~\ref{sec:photonics} we focus on the major photonic calibration sources used in precise astronomical spectrographs, including a description of the FFP and commercial illumination source used for this study. Section~\ref{sec:tests} describes the laboratory tests performed to gauge the stability of the device and estimate the maximum achievable velocity precision. Section~\ref{sec:apogee} briefly discusses the first light results on the APOGEE spectrograph. Finally, we present initial spectra of the FFP using the NIST 2-m Fourier transform spectrometer in Section~\ref{sec:NIST} .

\section{Summary of precise NIR/Optical wavelength references}
\label{sec:wvl_ref}

Connes et al. \cite{1994Ap&SS.212..357C} use a combination of a tunable Fabry-Perot and stabilized laser sources to measure stellar Doppler motion by varying the interferometer cavity to match the observed spectral shift. This technique, known as Absolute Astronomical Accelerometry, has demonstrated high precision ($\sim1 $ m s$^{-1}$) over short intervals on the varying solar spectrum but requires a vast array of fast, complex electronics to vary the interferometer cavity at the required precision and frequency.

McMillan et al.\cite{1994Ap&SS.212..271M} pass reflected sunlight through a fixed Fabry-Perot cavity and observe the changing slope of the spectral features transmitted through the interferometric filter to estimate the velocity shift over time. This method allowed for precise ($<4$ m s$^{-1}$) monitoring of solar RV variations, but requires a very bright target. Additionally, the Fabry-Perot inherently discards the majority of the stellar spectral features, significantly lowering the amount of Doppler information content. 

Wildi et al.\cite{2010SPIE.7735E.164W} designed and tested an optical FPI for the HARPS (High Accuracy Radial velocity Planet Searcher) spectrograph, finding 10 cm s$^{-1}$ precision over a 24 hour interval. However, optical path variations induced long term large scale velocity variations at the km s$^{-1}$ level. The variable optical path difference within the cavity introduced errors of up to 3 km s$^{-1}$ from night to night. This issue hinders high-precision velocity measurements over long periods, though these issues seem to have been solved in the most recent iteration of the device (see Wildi et al., these preceedings).

Uranium Neon (U/Ne) hollow-cathode lamps have been used to obtain RV precision of $<$10 m s$^{-1}$ in the Y and J-bands (900 - 1300 nm) using the Pathfinder instrument \cite{2010SPIE.7735E.231R} on integrated sunlight. This represents one of the few emission lamps calibrated specifically for use in NIR spectrographs for radial velocity measurements. A high density of lines is present in all commonly used NIR bands, making the lamp ideal for use as a precise reference source.

Mahadevan and Ge\cite{ 2009ApJ...692.1590M} explored a number of atomic combinations for NIR gas cells and concluded that the a combination of commercially available  H$^{13}$C$^{14}$N, $^{12}$C$_2$H$_2$, $^{12}$CO, and $^{13}$CO cells could conceivably provide a dense enough set of features for precise calibration in the H-band (see Figure~\ref{fig:cal_comp}), though no such cell has been deployed on a high-resolution astronomical spectrograph.

Bean et al.\cite{2010ApJ...713..410B} achieved $<$10 m s$^{-1}$ precision in the K-band using a NH$_3$ cell on the CRIRES instrument on the Very Large Telescope (VLT) on mid-M dwarfs, which are bright in these spectral windows. This presents an interesting wavelength region as K band is densely populated with telluric absorption features from H$_2$O and CH$_4$. The wavelength coverage of this method is rather narrow however, spanning roughly 36nm. This technique  required both the high resolution (R = 100,000) of CRIRES, and bright (K$<$8 mag) targets for high precision measurements. 

Blake et al. \cite{2010ApJ...723..684B} utilize the wealth of telluric absorption features near 2.3$\mu$m from H$_2$O and CH$_4$ to obtain RV precisions of $\sim$50 m s$^{-1}$ on ultracool dwarfs using the NIRSPEC\cite{1998SPIE.3354..566M} instrument on the Keck II telescope. There is a wide distribution of telluric lines through the NIR, including in the H-band. These lines are not static however, requiring unique reduction methods and careful atmospheric monitoring to reach sub 10 m s$^{-1}$ precision.

Anglada-Escud\'{e} et al.\cite{2012arXiv1205.1388A} explored using optimized gas cells ($^{13}$CH$_4$, $^{12}$CH$_3$D, and $^{14}$NH$_3$) for precise RV work in the K-band. $^{13}$CH$_4$ has an intrinsic spectrum nearly identical to $^{12}$CH$_4$, though the rovibrational mode differences between the two isotopes adds a slight offset between the two spectra. This offset allows for easy disentanglement of the gas cell reference lines from telluric  $^{12}$CH$_4$ lines. Passing starlight through the optimized $^{13}$CH$_4$ cell, they were able to achieve a preliminary RV precision of $\sim35$ m s$^{-1}$ on a bright reference star using just 6 nm of the K band on CSHELL (Cryogenic Echelle Spectrograph\cite{1994ExA.....3..309G}).

\section{Astrophotonic Calibrators}
\label{sec:photonics}

Fiber Bragg gratings have been proposed as accurate, low cost wavelength calibrators for astronomical spectrographs\cite{2003ASPC..298...93P}. This method consists of passing starlight through an optical fiber with an embedded Bragg grating structure that rejects a fixed selection of wavelengths. The advantage of this technique is that the reference requires no moving parts or power, leading to a simple addition to any fiber spectrograph that will yield several distinct reference absorption lines in a desired band. 

\begin{figure}
\begin{center}
\includegraphics[width=5.5in]{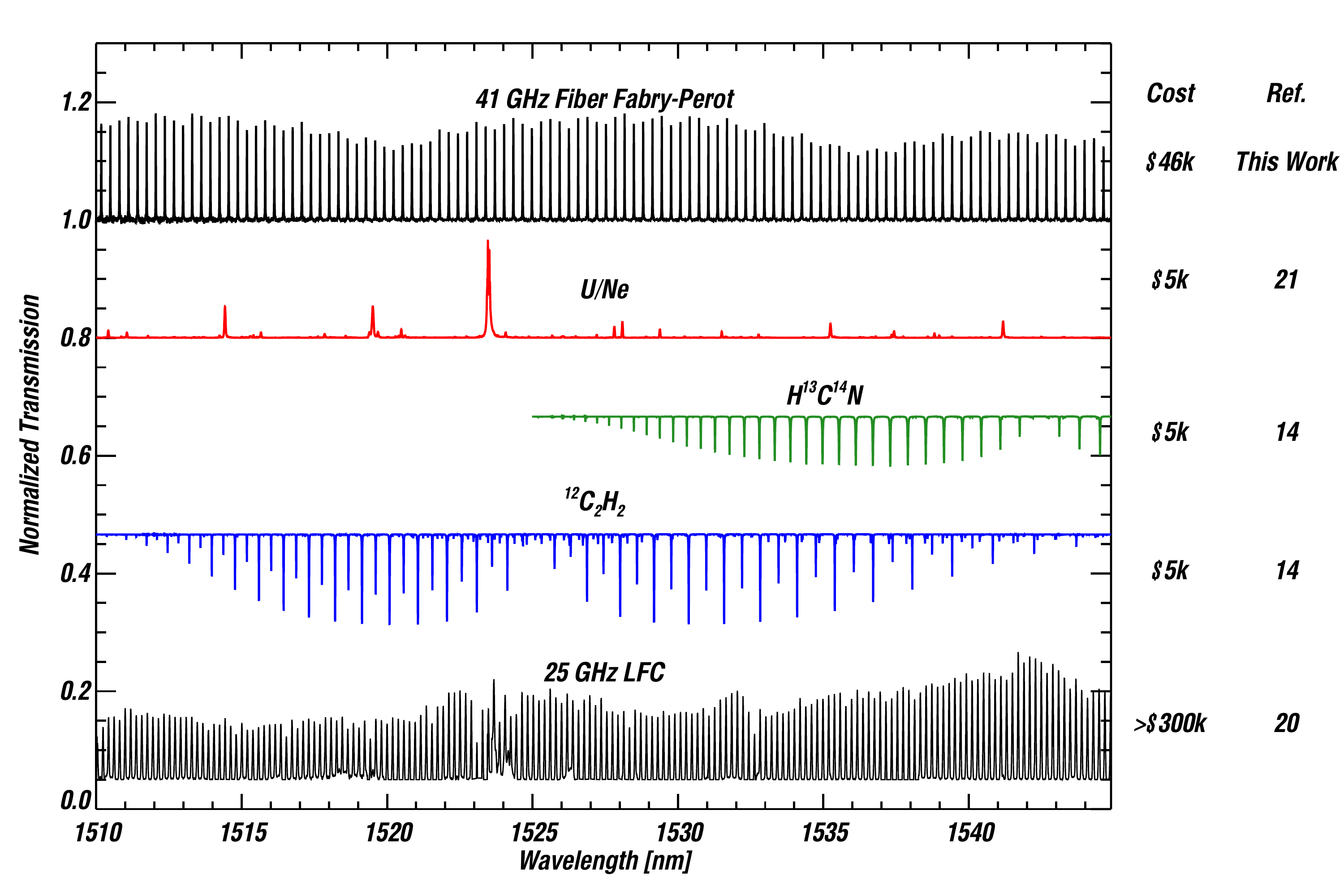}

\caption{Comparison of NIR wavelength calibration sources in the H-band, with approximate cost listed in the right column. Note the wide coverage of sharp FFP lines relative to the U/Ne lamp and absorption cells.  The FFP produces lines from 1260 - 1640nm. The FFP data is from a NIST FTS scan (this work, see Section~\ref{sec:NIST}), the U/Ne data is from a recently published high-resolution atlas\cite{2012ApJS..199....2R}, the gas cell data is from the NIST Standard Reference Materials database, and the LFC data was taken with the Pathfinder spectrograph\cite{2012OExpr..20.6631Y} at R$\sim$50,000.}

\label{fig:cal_comp}
\end{center}
\end{figure}

Laser frequency combs (LFC) have been tested in astronomical applications to success in
recent years. LFCs represent the pinnacle of
wavelength references, producing sharp emission features at known wavelengths and spacings. Despite the high cost and complexity in design, LFC development is a high priority in the spectroscopic community due to the extreme precision possibilities.
The difficulty lies in creating a custom NIR LFC that is stable over both long and short intervals. LFCs have
been shown to be stable to the cm s$^{-1}$ level\cite{2008Natur.452..610L}, but are not yet at the \lq{}turn-key\rq{} stage of use in astronomical instruments. An LFC spectrum taken using Pathfinder instrument\cite{2012OExpr..20.6631Y}  on the Hobby-Eberly Telescope is shown in Figure~\ref{fig:cal_comp}. This represents the first use of such a device on an NIR astronomical spectrograph, and we are pursuing further development of a stable LFC with our NIST colleagues.

\begin{figure}  
\begin{center}
\includegraphics[width=3.3in]{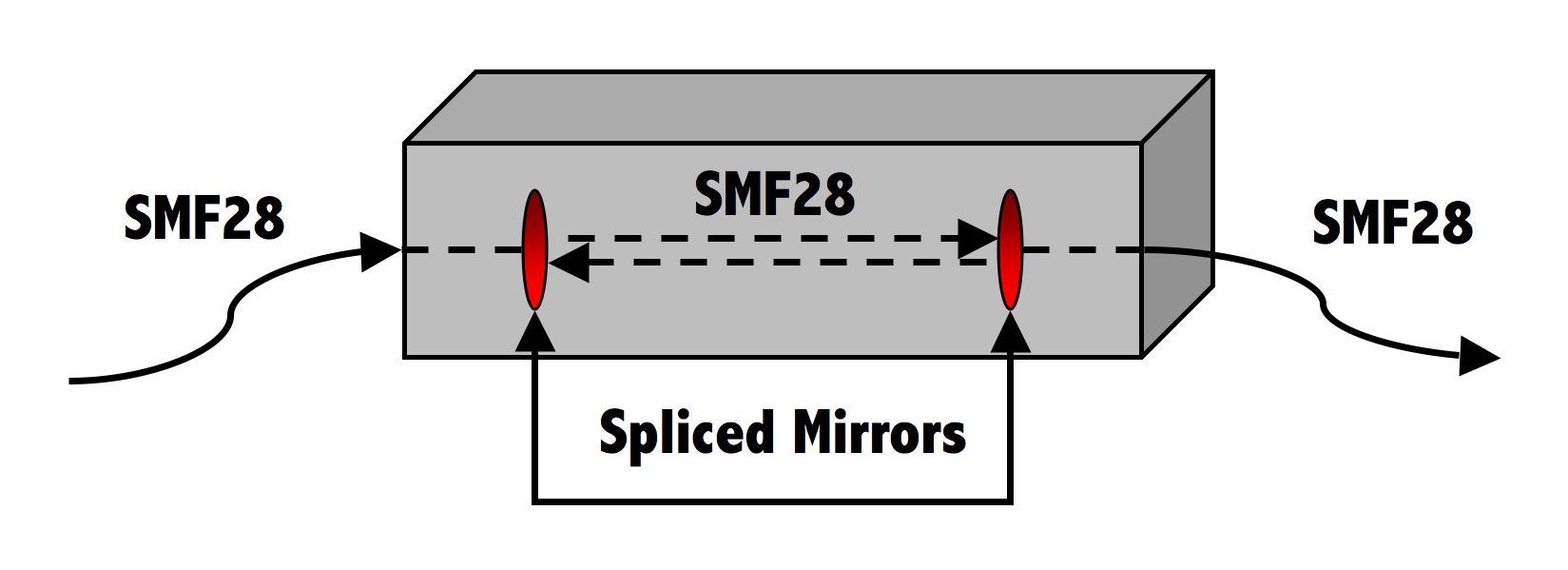}\includegraphics[width=3.3in]{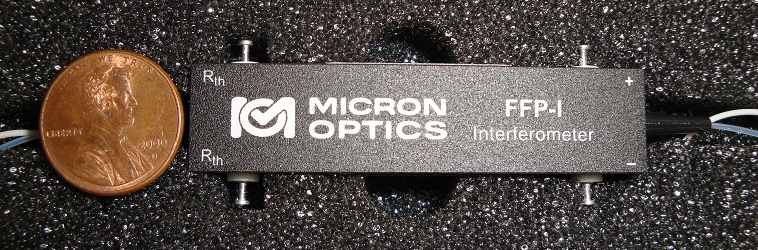}
\caption{Left: Fiber-FP cavity schematic. The device consists of SMF-28 fiber and custom coated reflective mirrors. Right: Picture of Micron-Optics H-band FFP. The pins on either side of the casing are leads for the temperature control system.}
\label{fig:ffp}
\end{center}
\end{figure}

The FFP used in this study was purchased from Micron Optics\footnotemark[1] and consists of SMF-28 fiber with two dielectric stack mirrors spliced directly into the fibers (see Figure~\ref{fig:ffp}). The chosen mirror coatings and cavity spacing yield a finesse of 36 and a free spectral range (FSR) of 41 GHz\footnote{41 GHz corresponds to 0.33 nm spacing at 1550 nm.}. The combination of chosen finesse and FSR values for our device is not standard, so custom mirror coatings were requested that required a longer ($\sim$6 month) lead-time. The device also has a built in 10 k$\Omega$ thermistor and $2$ A thermo-electric cooler for precise temperature control. 
The uniform density of emission lines produced by the FFP gives a much wider bandwidth over which RV measurements are possible. 
The FFP provides a wealth of sharp, uniformly spaced emission lines from 1260 to 1640 nm. The interferometric nature of the FPI does not by itself yield an absolute wavelength reference, and a stabilized laser or emission lamp reference source is needed for absolute wavelength calibration.  This could conceivably be done by observing both a U/Ne lamp and the FFP, either simultaneously through separate fibers or in succession. The U/Ne would provide absolute tracking of the FFP lines, and the FFP would provide broad-band calibration for the spectrograph. This combination will be explored in the near future as an absolute wavelength reference. 

The device requires a specialized light source for operation, as coupling a significant amount of light through the $10\mu$m single-mode fiber (SMF) FFP  is not achievable using common broadband lamp sources. The continuum source chosen is a white light fiber laser source, discussed in Section~\ref{sec:superK}.

Light reflecting off the dielectric mirror stack is susceptible to wavelength-dependent phase dispersion, which may alter the intensity or central wavelength of a given emission feature. The extent to which this affects the transmission spectrum is not yet known, but if the dispersion is temporally stable it will not greatly affect the long term velocity stability.

\subsection{Expected Output}
The theoretical FFP output spectrum can be quantified as a simple interferometric airy function:
\begin{equation}
I(\delta) = \dfrac{1}{1 + \mathcal{F}\sin{\left(\dfrac{\delta}{2}\right)}^2},
\end{equation}

where $\delta$ is the cavity delay, related to the free spectral range (FSR) and wavenumber $k$: $\delta = \frac{ck}{\mathrm{FSR}}$. $ \mathcal{F} $ is the finesse coefficient, related to the standard finesse value $F$ through:

\begin{equation}
\mathcal{F} = \left(\dfrac{2F}{\pi}\right)^2.
\end{equation}

Few wavelength references yield the ability to model the inherent emission spectrum using basic laws of interference. This ability to model the intrinsic emission line profile can allow for precise instrument profile modeling, though mirror phase dispersion and collimation issues can complicate the inherent functional form. An advantage of the Micron Optics\footnotemark[1] fiber FPs that the measured transmission profile is very close to the ideal airy profile.

A common metric for estimating radial velocity information content from spectra is the quality factor Q\cite{2001A&A...374..733B}. The Q factor represents a quantitative assessment of the quality and richness of the lines within a spectrum, or the amount of extractable Doppler information given an intrinsic emission spectrum at a given instrument spectral resolution. It is also independent of spectral flux.

The functional form of Q for a given spectrum I$_\lambda$ spanning N pixels is given by:

\begin{equation}
Q = \frac{\sqrt{{\sum\limits_{i}^N W_i}}}{\sqrt{{\sum\limits_{i}^N}{I_\lambda}(\lambda_i)}},
\end{equation}
where I$_\lambda$($\lambda_i$) is the intensity of the intrinsic spectrum at a given pixel $i$, $\lambda_i$ is the wavelength of that pixel, and W$_i$ is the optimum weight, given as (neglecting detector noise):

\begin{equation}
W_i = \frac{{\lambda^2_i}({\partial}{I_\lambda}(\lambda_i) /{\partial}\lambda_i )^2}{{I_\lambda}(\lambda_i)}.
\end{equation}

\begin{figure}
\begin{center}
\includegraphics[width=4.0in]{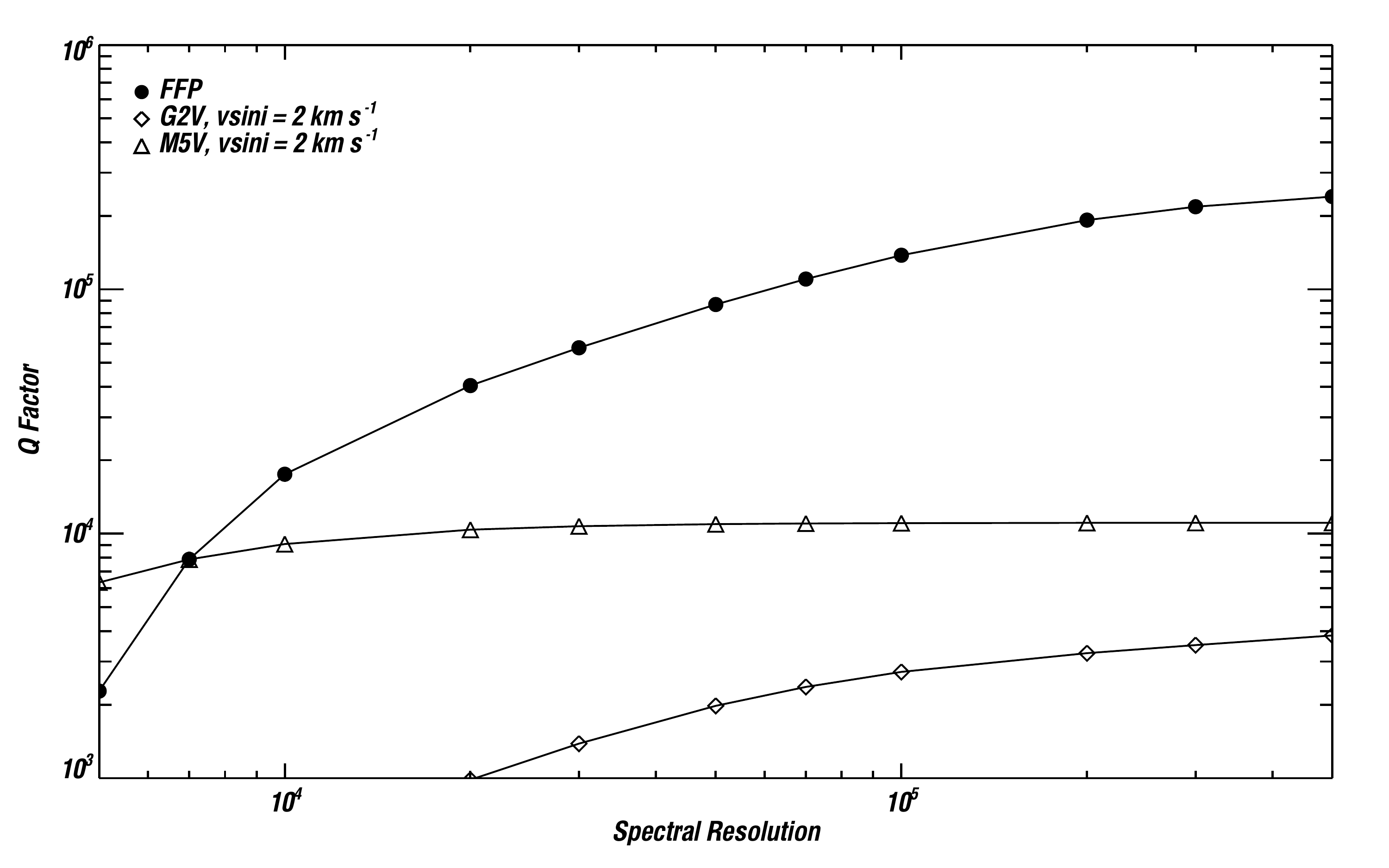}
\caption{Q factor for the theoretical FFP spectrum from 1200 nm to 1700 nm as a function of spectrograph resolution. The quality factor for model M5V spectrum and archived solar spectrum are over-plotted for comparison. The amount of velocity information in the FFP emission lines is much higher than in the stellar spectra due to the wide coverage of sharp lines.}
\label{fig:ffp_qfactor}
\end{center}
\end{figure}

The quality factor for a model FFP spectrum from 1200 to 1700 nm is shown in Figure~\ref{fig:ffp_qfactor} as a function of spectral resolutions. 
The overall velocity uncertainty for a spectral range can be determined through the quality factor:

\begin{equation}
\sigma_{\mathrm{RV}} = \frac{c}{Q\sqrt{N_{e^-}}},
\end{equation}
where $N_{e^-}$ is the total number of detected photoelectrons in the spectrum. The high quality factor of the FFP, combined with the unresolvable lines and bright illumination source, imply the spectrum is ideal for use as a stable and precise wavelength reference for a wide range of spectrograph resolutions.

\subsection{Expected Stability}
\label{sec:stability}
 The simple nature of the FFP design reduces the number of physical parameters that introduce velocity measurement error. The interferometer mirror spacing (L) and refractive index (n) of the SMF-28 fiber are both sensitive to temperature fluctuations that cause the interferometric emission line output to drift. These spectral shifts can mimic velocity shifts, thereby introducing measurement error. The phase change of the output pattern d$\phi$/dT is related to the cavity width and refractive index through:

\begin{equation}
\phi = {\alpha}nL,
\end{equation}
where $\alpha$ is an unknown constant. The fraction phase change as a function of n and L can be described as:
\begin{equation}
\frac{\mathrm{d}\phi}{\phi} = \frac{\mathrm{d}L}{L} + \frac{\mathrm{d}n}{n}.
\end{equation}

This corresponds to a temperature dependent wavelength shift of:

\begin{equation}
\frac{\mathrm{d}\lambda}{\mathrm{d}T} = {\lambda_o}\left(\frac{1}{L}\frac{\mathrm{d}L}{\mathrm{d}T} + \frac{1}{n}\frac{\mathrm{d}n}{\mathrm{d}T}\right).
\end{equation}

For a standard SMF-28 fiber, the physical coefficient values are listed in Table~\ref{tab:ffp_params}. 

\begin{table}
\begin{center}

\caption{Wavelength coverage and physical coefficients of FFP at 1550 nm, 298 K}

\begin{tabular}{ccc}

\hline
\hline
Parameter & Value & Reference \\
\hline

Operating wavelength range (nm) & 1260 - 1640 &	Micron Optics\footnotemark[1] 	\\

Finesse & 36 &	Micron Optics\footnotemark[1] 	\\

Free spectral range  (nm) &  0.33	& Micron Optics\footnotemark[1] 	\\

\ \ \ \ \ \ \ \ \ \ \ \ \ \ \ \ \ \ \ \ \ \ \ \ \ \ \ (GHz)	&	41	&	\\

n 	&	1.44 		&	\cite{1267474}   \\
$\dfrac{\mathrm{d}n}{\mathrm{d}T}$ (K$^{-1}$)		&	$1.06323\times10^{-5}$	&	\cite{1267474}   \\

$\dfrac{1}{L}\dfrac{\mathrm{d}L}{\mathrm{d}T}$	 (K$^{-1}$)&	$5.5\times10^{-7}$	&	\cite{1267474}   \\

\hline
\label{tab:ffp_params}
\end{tabular}
\end{center}

\end{table}

We estimate the velocity stability of the device to be $\frac{\mathrm{dv}}{\mathrm{dT}} $= 2.2 km s$^{-1}$ K$^{-1}$ at a wavelength of 1550 nm.
Section 5 shows achieved control stability using the internal FFP temperature electronics with
a bench-top temperature controller. The peak-to-peak temperature variation over a 24 hour period is 0.15 mK, meaning the
expected velocity stability based on temperature fluctuations alone is 0.35 m s$^{-1}$. Note that this does not include any variation due to phase dispersion from the dielectric mirror stacks, though we expect this effect to be minimal relative to the temperature sensitivity. 

\subsection{Illumination Source}
\label{sec:superK}
The single-mode fibers used in the FFP reduce the number of continuum sources that have sufficient broad-band intensity to illuminate the emission lines. High power broad-band lamps do not have the required coupling efficiency to yield a practical number of photons per FFP line in reasonable integration times. 
We tested a broadband, 20 W Quartz Tungsten Halogen (QTH) light source optimized for multi-mode fiber delivery, but found the light loss in coupling to a SMF to be too great. Integration times needed to achieve high S/N ($>100$) spectra were not practical with this source. 

Super-luminescent diodes (SLEDs) can provide several mW of power through SMF fibers in the NIR, but are generally limited to narrow (50 - 100 nm) bandwidths. A combination of several SLEDs would be required to provide full coverage of the H-band.

\begin{figure}  
\begin{center}
\includegraphics[width=3.3in]{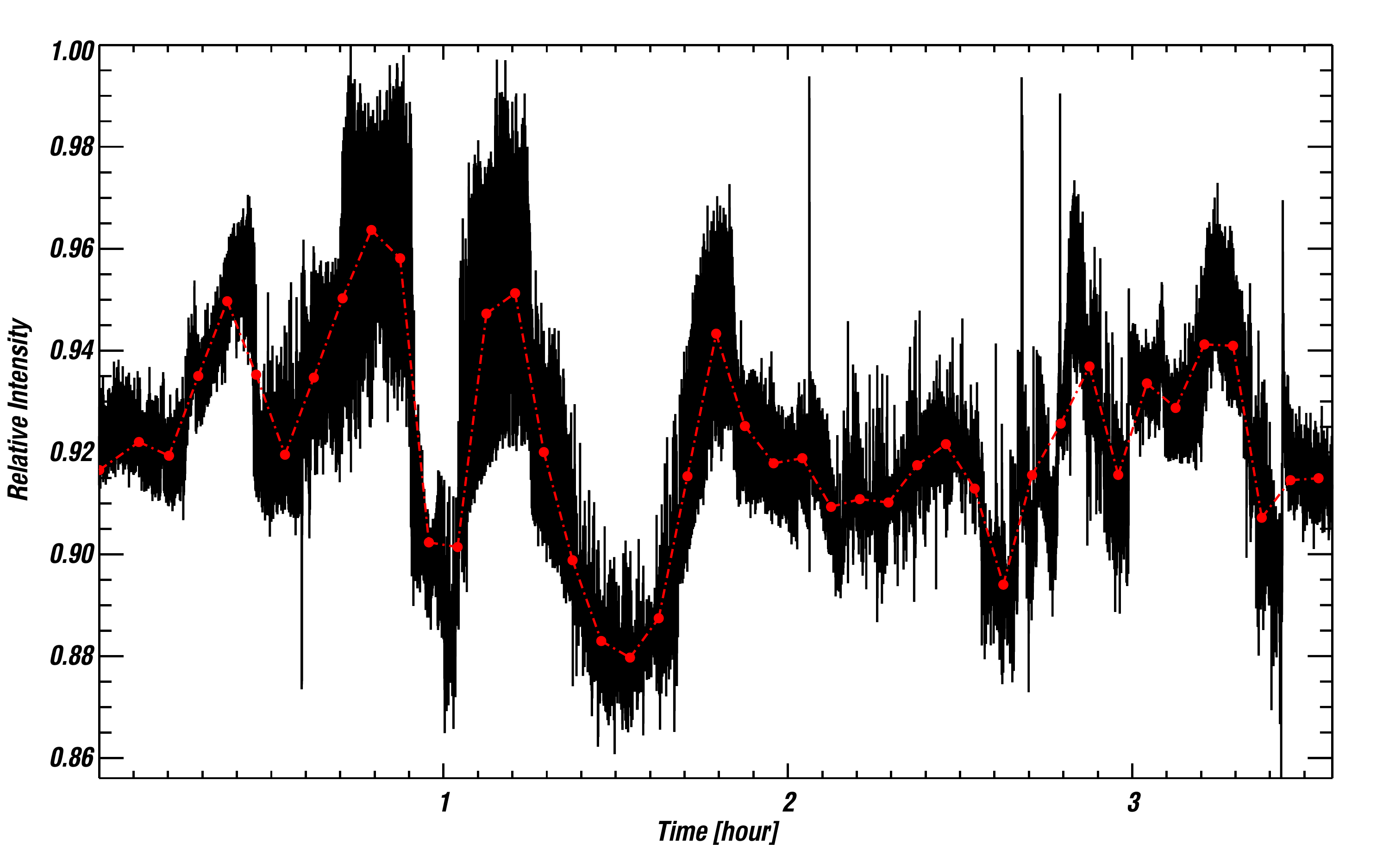}\includegraphics[width=3.3in]{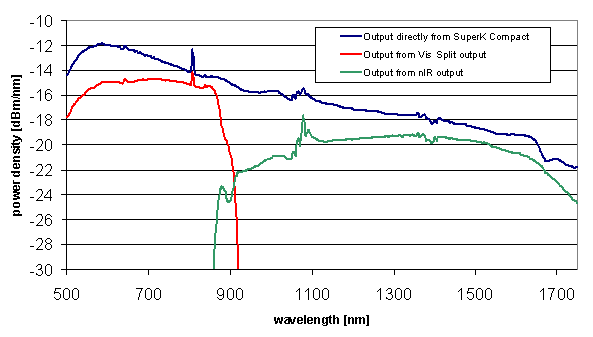}
\caption{Left: Portion of intensity time series of SuperK supercontinuum source. The red line is the data binned into 5 minute intervals. Right: Output power spectrum of SCS, with both VIS and NIR components.}
\label{fig:superk}
\end{center}
\end{figure}

The illumination source chosen is a broadband (500 - 2500 nm) supercontinuum source (SCS) that outputs 100 mW integrated across all wavelengths. The unit is an off-the-shelf source developed by SuperK Photonics\footnotemark[1]. The laser uses standard SMF delivery, providing a bright power source directly through the FFP at a pulsation rate of 24 kHz. An optical splitter is used to separate the visible and NIR regions into separate fibers (see Figure~\ref{fig:superk}). An image of the SuperK Compact\footnotemark[1] source and output spectrum are shown in Figure~\ref{fig:superk}.

The SCS is one of the few available broadband sources capable of achieving significant output through the FFP. The output raddiance is several orders of magnitude higher than common QTH lamps. This high power concentrated within a fiber is required for the FFP system, as the device is an inherent filter of $>$95\% of incoming light. Furthermore, the addition of necessary optical instruments that reduce overall throughput along the fiber path (such as integrating spheres) will further decrease the intensity of the FFP lines.  The stability of the SCS is quoted to be 0.3dB ($\sim$7\%). We measure the output stability to be roughly consistent with this value using a Ge diode and power meter (see Figure~\ref{fig:superk}).

\section{Laboratory Tests}
\label{sec:tests}
The major components of the FFP system were carefully selected to maximize instrument efficiency and temperature control. The areas of concern were precise temperature control, high intensity fiber fed illumination source, and interferometer mirror coatings. Finesse and free spectral range were chosen for optimum line spacing and throughput in the NIR. A bench-top Stanford Research Systems\footnotemark[1] temperature controller was used for fine control of the FFP temperature. The controller is capable of sub-mK temperature control for precise coolers in good thermal contact. 
\subsection{Temperature Stability}

We expect temperature variations in the FFP to be the dominant contribution to the velocity error. Therefore, the temperature control precision of the FFP will determine the overall stability of the interferometer cavity. 

\begin{figure}  
\begin{center}
\includegraphics[width=3.3in]{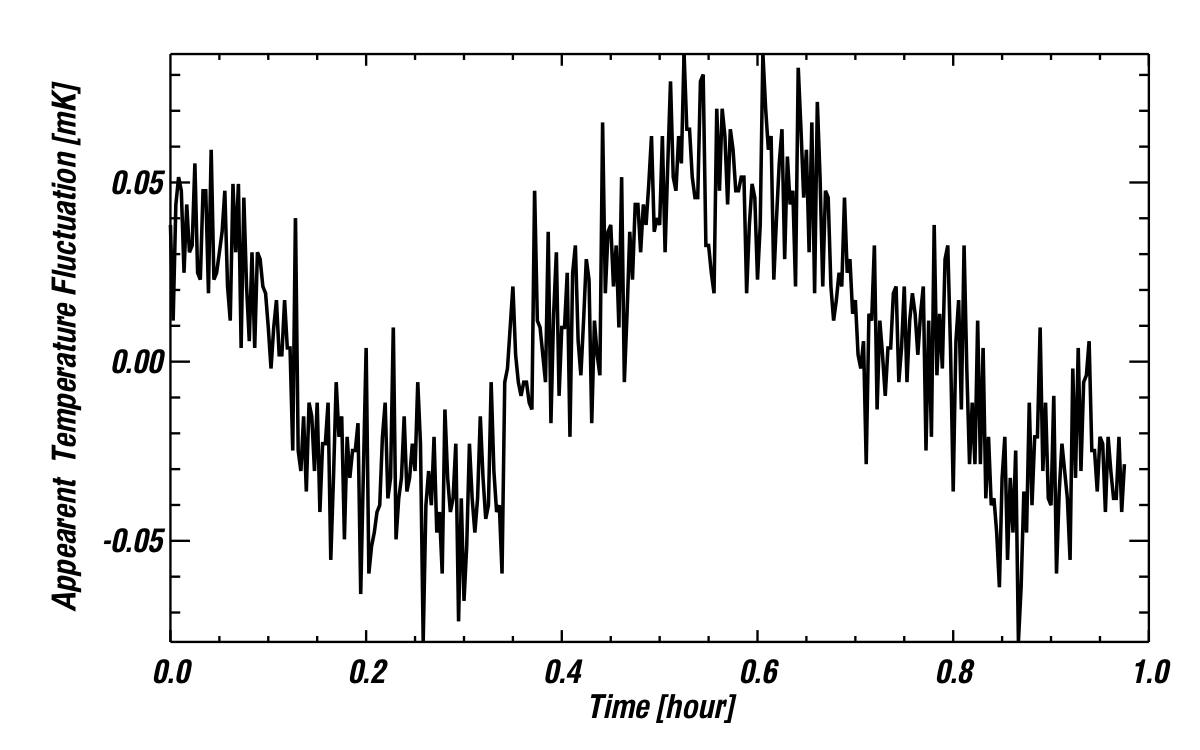}\includegraphics[width=3.3in]{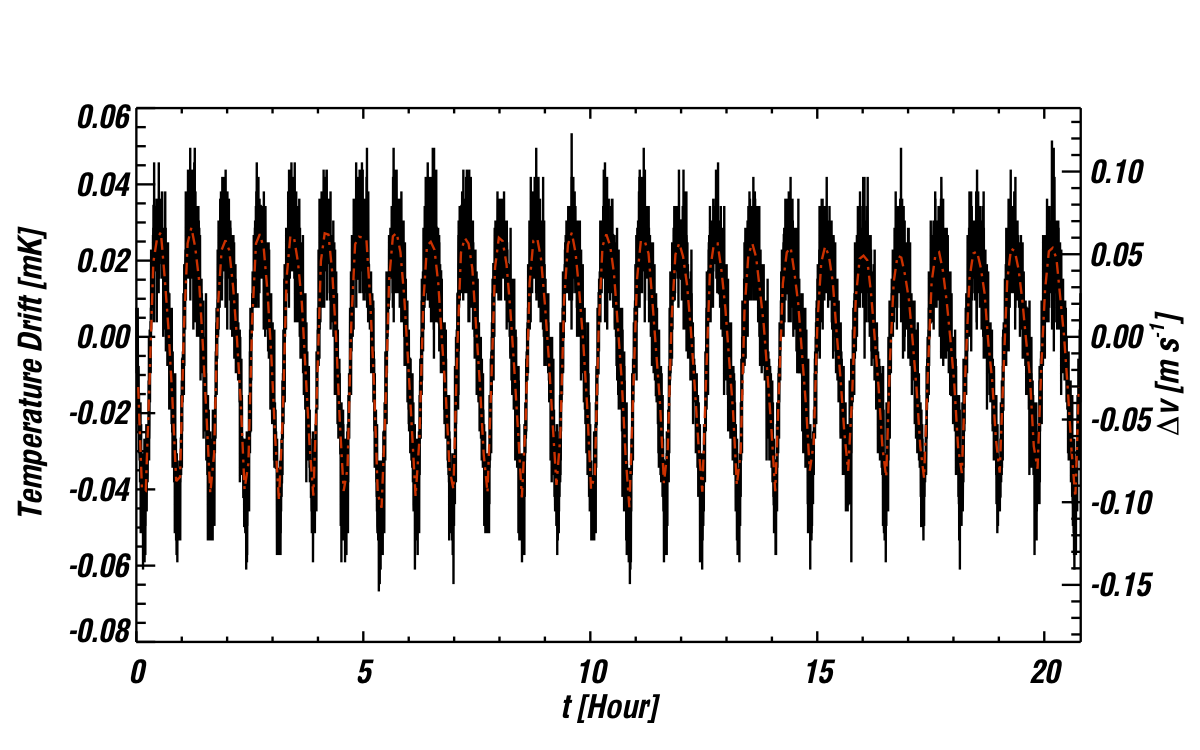}
\caption{Left: Temperature reading of precision resistor as measured by the temperature controller. The peak-to-peak variation in measurements yield an estimate for maximum achievable temperature reading precision. Right: Section of temperature (with theoretical velocity drift at 1550 nm) time series of internal FFP Thermal-Electric Cooler (TEC) unit. The TEC was set to 298 K for a 24 hour period. The red line is data at 5 minute binning, indicative of a typical bright observational exposure.}
\label{fig:ffp_temp}
\end{center}
\end{figure}

To estimate the achievable control precision of the temperature controller, we first connected a Vishay\footnotemark[1] high precision resistor to the temperature control system to emulate the resistance readout of the FFP thermistor. The resistor has a value of $98988\pm0.5$ $\Omega$, similar to the to 10 k$\Omega$ thermistor contained in the FFP. Unlike the thermistor, the thermal response of the precision resistor is typically 5$\times 10^{-4}$ $\Omega /$ K near 298 K, meaning any practical temperature changes in the resistor ($<10$ K) will yield a negligible resistance change read by the controller. With this \lq{}fixed temperature\rq{} source soldered directly to the temperature reading card of the controller, a baseline reading precision of the temperature controller can be estimated. The results of this test are depicted in Figure~\ref{fig:ffp_temp}. We estimate the peak-to-peak variations of the temperature reading system to be $\sim$0.13 mK over a 24 hour period. This gives an estimate of the electronics noise limit of the temperature control capability for the system.

Secondly, we replicated the temperature electronics within the FFP unit using an identical Quality Thermistors\footnotemark[1] thermistor and Laird\footnotemark[1] TEC unit. Placing these two devices in close thermal contact in an insulated fiber-glass enclosure in a temperature regulated laboratory, we were able to hold the temperature of the system to better than 0.15 mK at 298 K using the high precision temperature controller. 

Finally, the temperature feedback loop parameters in the temperature controller were optimized for the thermal response of the FFP and the ambient temperature fluctuations and the FFP was placed inside thermally insulated foam wrap for 24 hours. Setting the temperature to 298 K, the absolute peak-to-peak temperature variations of the cavity were measured to be 0.15 mK (Figure~\ref{fig:ffp_temp}). This high precision control, nearly at the electronics limit of the controller, demonstrates the exquisite thermal contact between the internal TEC and thermistor, as well as the fast thermal response of the FFP as a whole. This peak-to-peak variation corresponds to a velocity error contribution of 35 cm s$^{-1}$ due to temperature fluctuations alone, though this cannot currently be measured using any available NIR grating spectrograph. 

\subsection{Narrow-band Laser Scan}

\begin{figure}[!h]
\begin{center}
\includegraphics[width=4.0in]{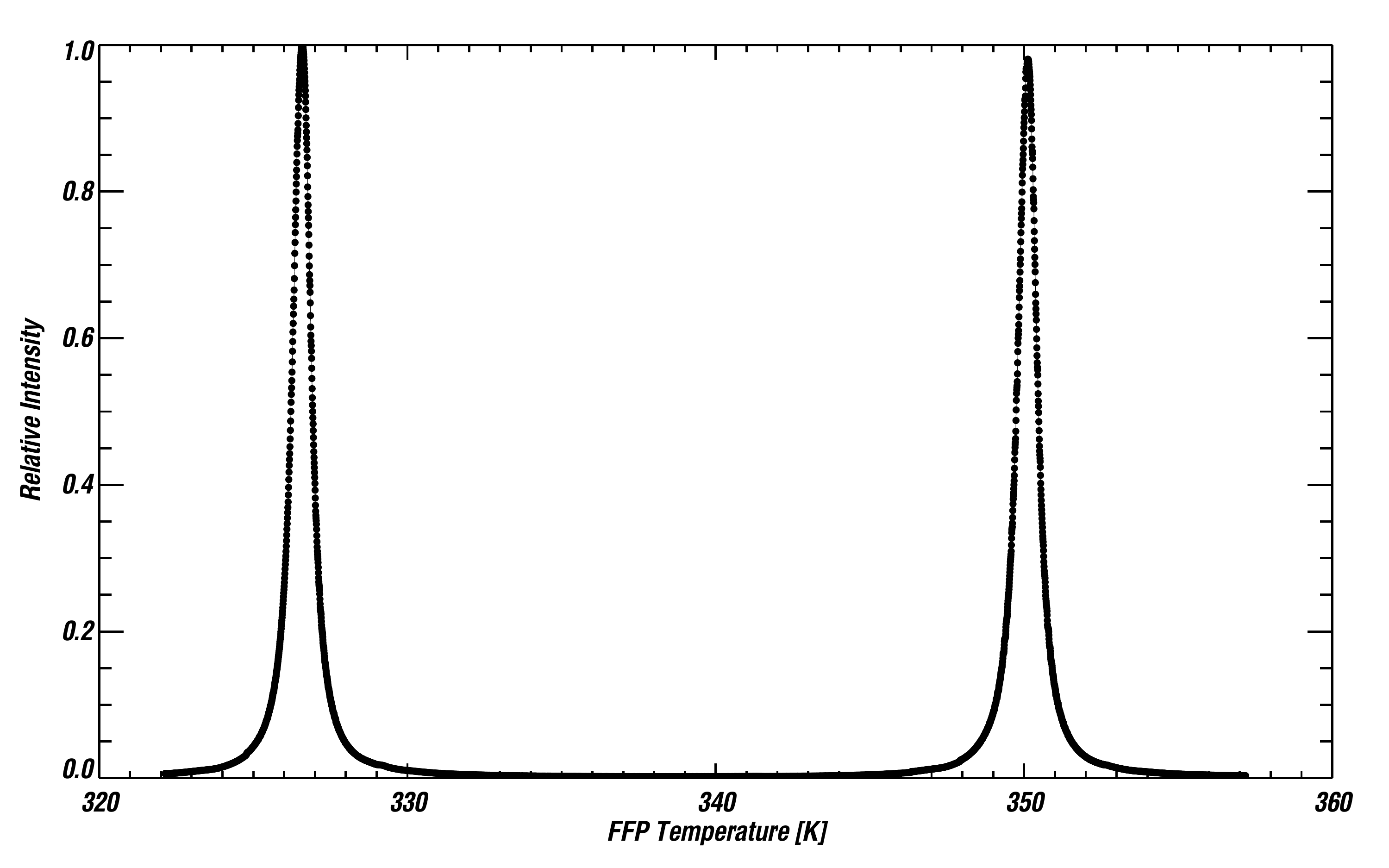}
\caption{38 degree scan using narrowband (10 MHz) 1550.6 nm laser source.}
\label{fig:ffp_spec_stable}
\end{center}
\end{figure}

The temperature sensitivity of the FFP spectrum allows for precise, high resolution sampling of the transmission line profiles.
A narrowband Acronym Fiber Optics\footnotemark[1] 1550.6 nm distributed feedback (DFB) laser was used to illuminate the FFP while the temperature of the FFP was increased so that the cavity scanned through two interferometric peaks. The FP interferometer acts a natural narrow filter for off-line wavelengths, allowing for a precise line profile to be measured. The laser source was held at fixed temperature and wavelength while the transmitted intensity through the FFP transmission lines was monitored using a NIR Ge diode.  The final laser scan is shown in Figure~\ref{fig:ffp_spec_stable}. The peak wavelength spacing to temperature spacing is in excellent agreement with the theoretical value at this wavelength based on the spacing to temperature relation derived in Section~\ref{sec:stability}.

\section{Commissioning on APOGEE}
\label{sec:apogee}

\begin{figure}
\begin{center}
\includegraphics[width=6.5in]{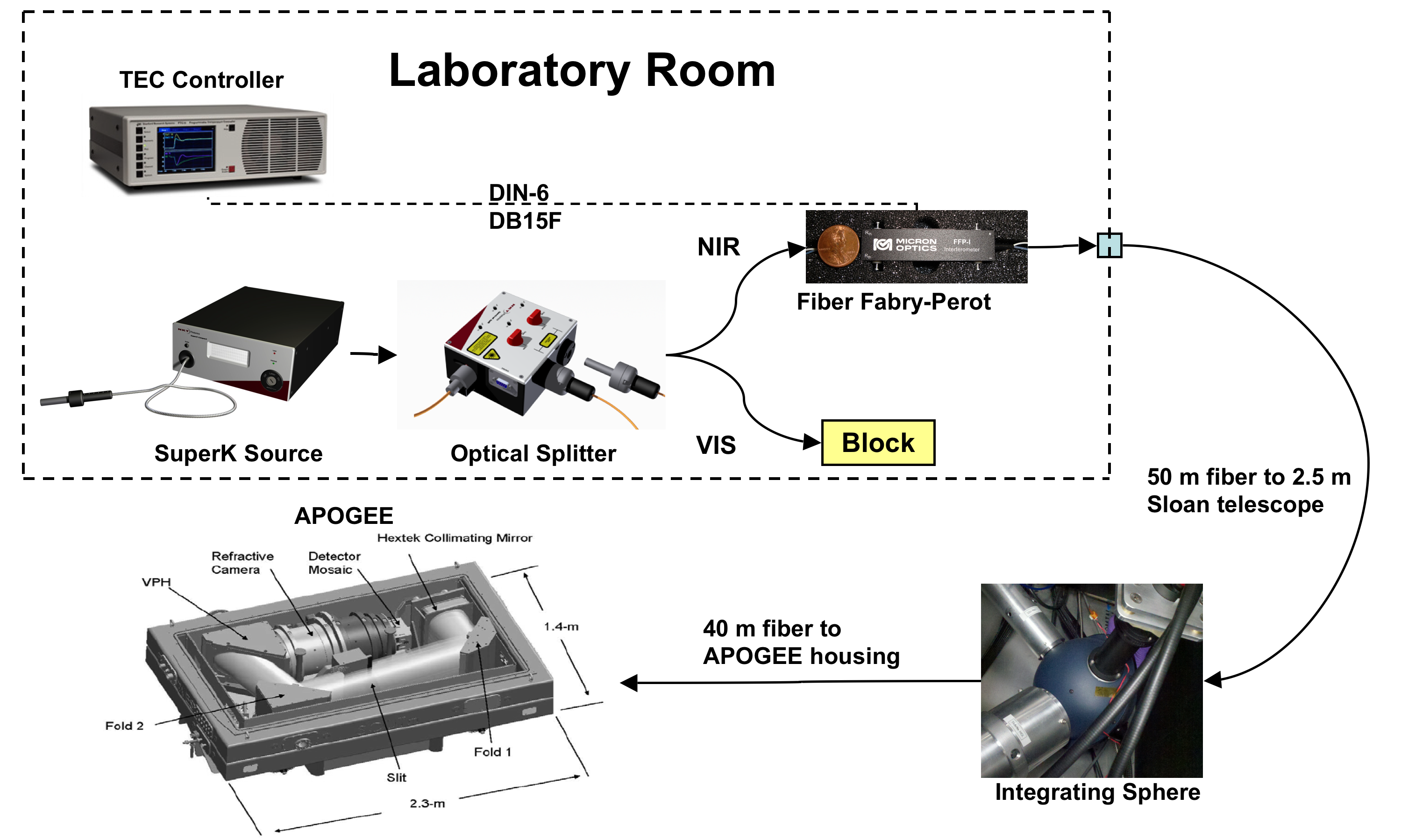}
\caption{Setup of the FFP system on APOGEE. The FFP and supporting equipment were placed in a building adjacent to the 2.5 m Sloan telescope housing. A 50 m fiber was fed to the integrating sphere (located within the dome), which is attached to the APOGEE spectrograph\cite{2010SPIE.7735E..46W}.}
\label{fig:ap_setup}
\end{center}
\end{figure}

The Apache Point Observatory Galactic Evolution Experiment (APOGEE), as part of the Sloan Digital Sky Survey III\cite{2011AJ....142...72E} (SDSS-III), is surveying 100,000 red stars spanning a large range of galactic latitudes to study the chemical and kinematical history of all Milky Way populations.  APOGEE uses a recently commissioned 300-fiber R$\sim$22,500 H-band fiber-fed spectrograph\cite{2010SPIE.7735E..46W}.  Such an instrument is a prime candidate for the addition of a stable, precise wavelength reference source to augment its existing calibration toolbox of traditional Th/Ar and U/Ne Hollow Cathode Lamps and enable high radial velocity precision.  One of the APOGEE ancillary science programs includes monitoring several hundred M-stars throughout the Galactic plane.  For instruments with these resolutions, the output of a high finesse ($>$30) FPI will be nearly indistinguishable from an LFC with similar emission line spacing. We estimate the intrinsic FFP line widths to yield emission lines with widths 6-10\% larger than the instrument PSF. A block diagram of the FFP setup on APOGEE is shown in Figure~\ref{fig:ap_setup}.

\begin{figure}
\begin{center}
\includegraphics[width=5.5in]{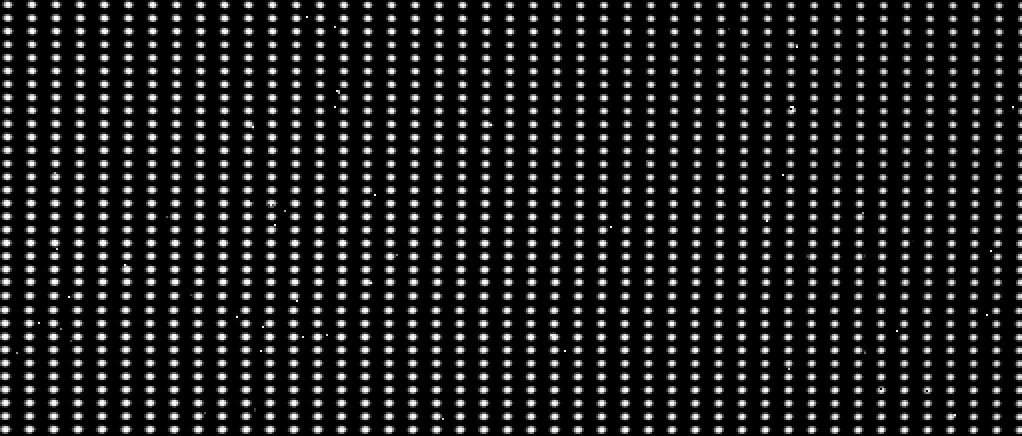}
\caption{Portion of first light APOGEE FFP frame. Each row of emission lines is an individual fiber spectrum. Approximately 120,000 inteferometric emission peaks cover the 300 fibers across the APOGEE detector mosaic.}
\label{fig:firstlight}
\end{center}
\end{figure}

The H-band FFP achieved first light on APOGEE in Spring 2012. A portion of the first light image is shown in Figure~\ref{fig:firstlight}. The device allowed for precise characterization of many spectrograph properties, including instrument drift in both spatial and spectral directions, detector persistence, wavelength solution stability, and PSF variations. A more thorough discussion of the analysis will be presented in a separate publication. 

\section{NIST FTS Scan}
\label{sec:NIST}

A high resolution (R$\sim$750,000) scan of the FFP was taken using the NIST 2-m Fourier transform spectrometer (FTS). The goal of this scan was to measure the absolute wavelengths of the FFP emission lines to high accuracy (15 - 20 m s$^{-1}$) by calibrating against a standard C$_2$H$_2$ reference cell. This would allow for an accurate determination of both line centers and temperature-to-wavelength conversion. 

The initial spectra taken using the SCS and FFP were unexpectedly noisy (S/N $<50$), even after combining a large number of coadded frames. This high noise level is attributed to the unstable pulsation rate of the Q-switching laser that is contained in the SuperK\footnotemark[1] SCS.   We observed the repetition rate of the source to vary by between 5 \% and 10 \% around the 24 kHz frequency. Since the NIST 2-m FTS takes data at equal time intervals, fluctuations in the frequency of the super-continuum source add noise to the interferogram and resulting spectrum. This prevented us from getting a high SNR spectrum using the SuperK source.

\begin{figure}
\begin{center}
\includegraphics[width=5.5in]{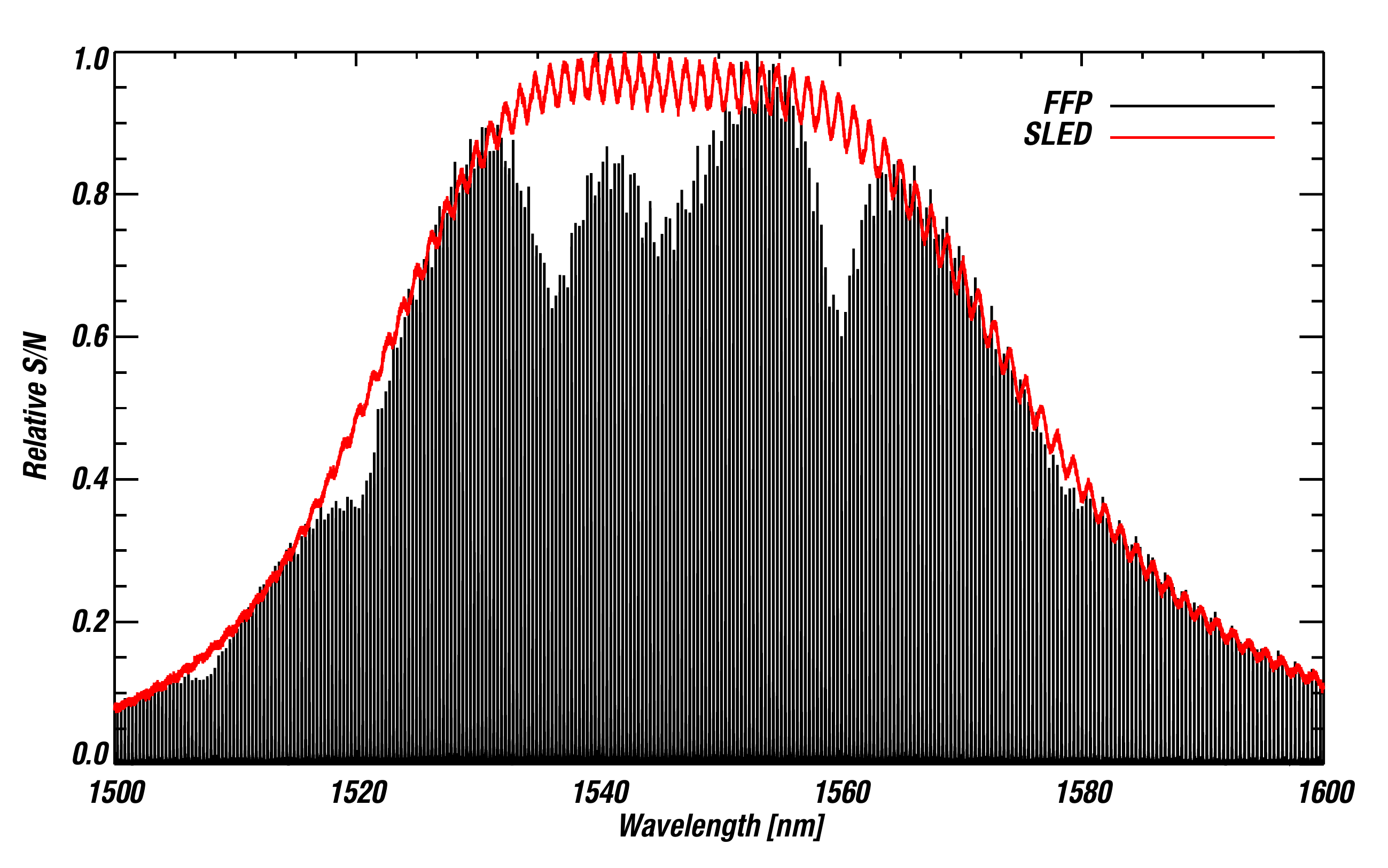}
\caption{FTS spectrum of FFP illuminated with a commercial NIR SLED source. Note the periodic ripples in the SLED spectrum and the FFP emission peak intensity deviation from the continuum at certain wavelengths.}
\label{fig:ffp_sled}
\end{center}
\end{figure}

We instead opted to use a more narrow-band 1550 nm Superluminescent light emitting diode (SLED) source to illuminate the FFP. The SLED produced $\sim5$ mW of power across 1500 - 1600 nm, sufficient for scans of the FFP at S/N $\sim400$ with $\sim$30 coadds. An example FTS scan across the operating wavelengths of the SLED is shown in Figure~\ref{fig:ffp_sled}.  

\begin{figure}
\begin{center}
\includegraphics[width=5.5in]{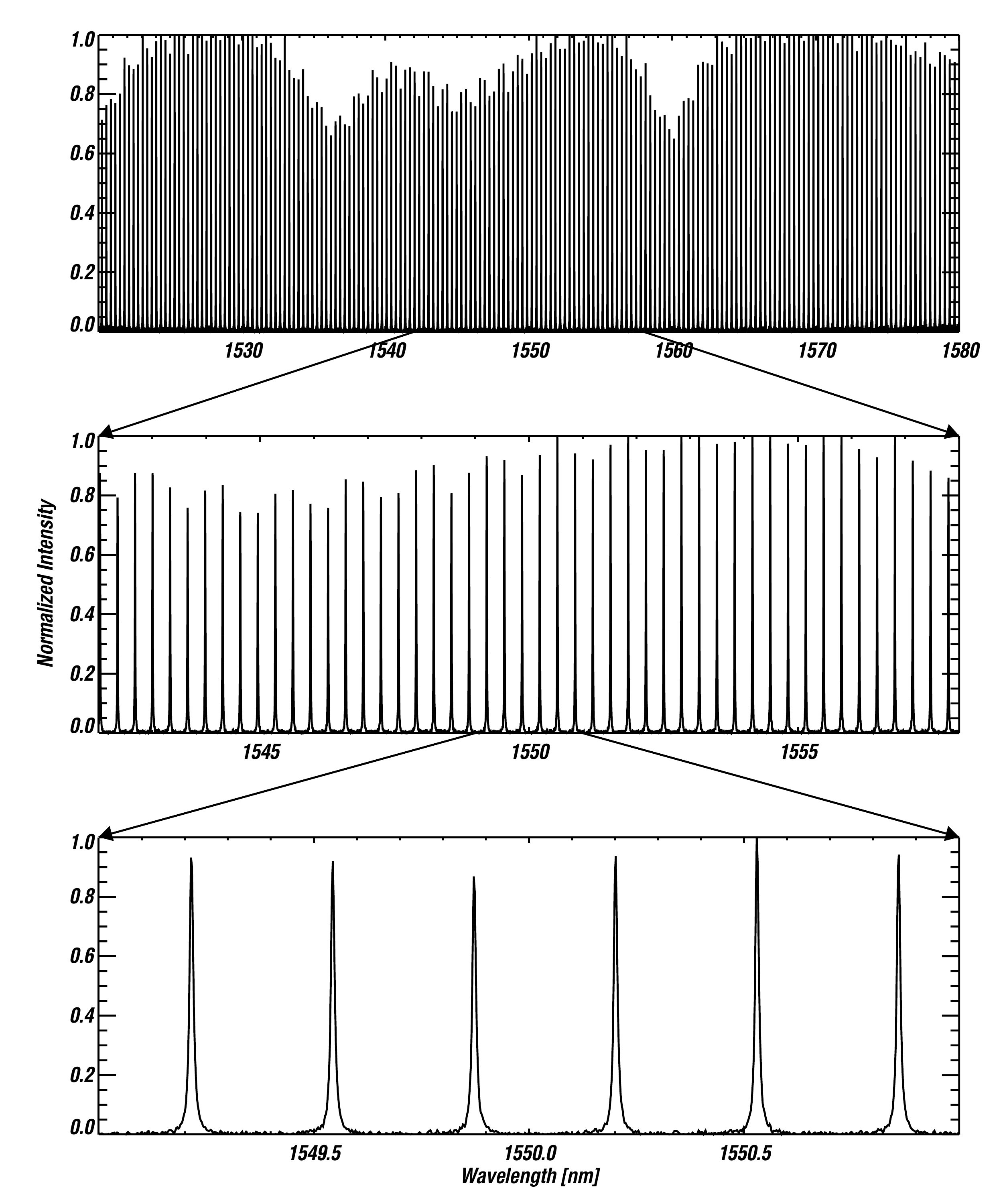}
\caption{NIST Fourier transform spectrum of the FFP using H-band SLED source at narrowing wavelength intervals.}
\label{fig:nist_scan}
\end{center}
\end{figure}

\begin{figure}
\begin{center}
\vspace{-40pt}

\includegraphics[width=6.5in]{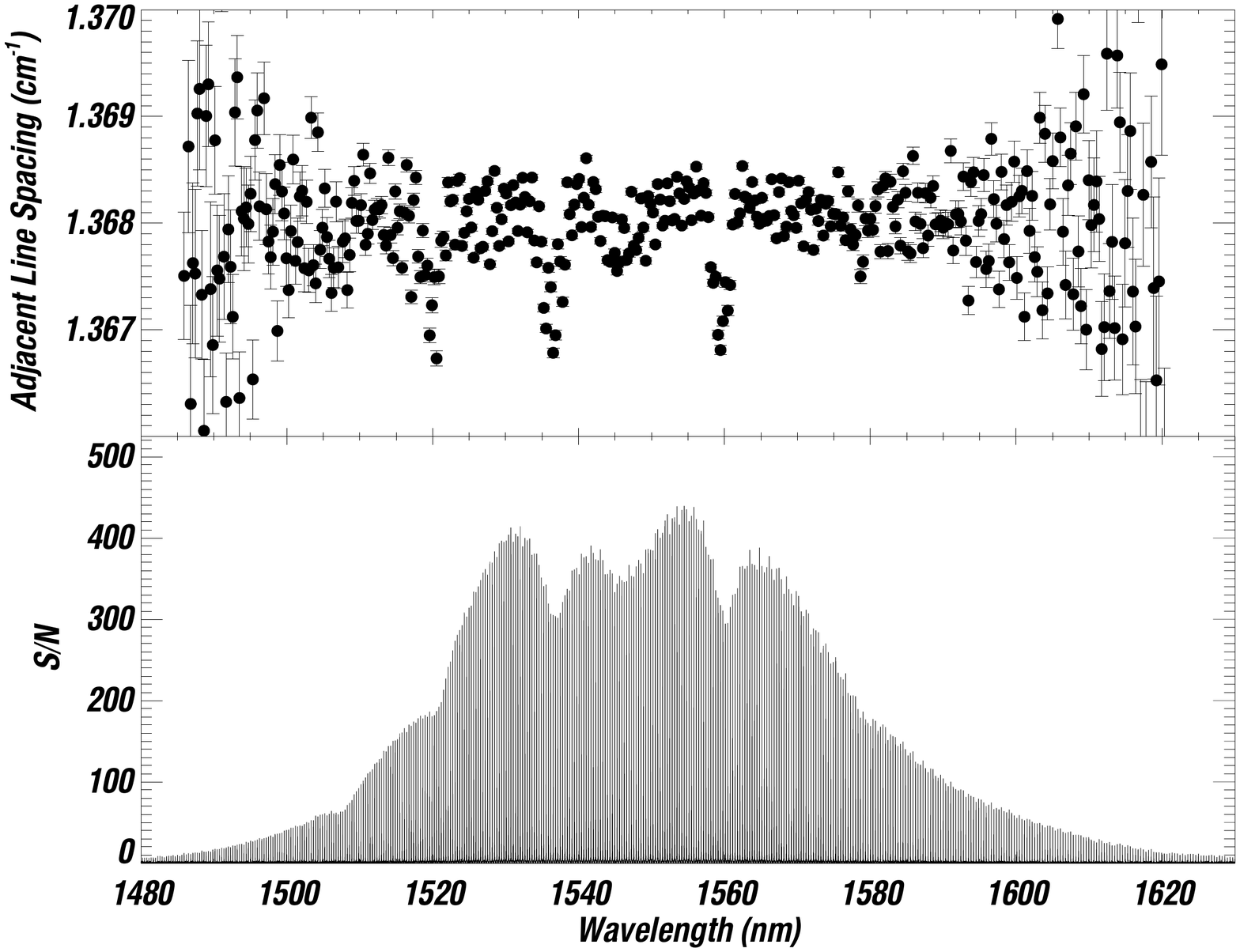}
\vspace{-20pt}
\caption{Measured line spacings between adjacent emission lines in the FFP spectrum for two FTS scans. Note the larger deviations in regions where the FFP peak intensities deviate from the SLED continuum, possibly due to phase dispersion at those wavelengths.}
\label{fig:line_spacing_err}
\end{center}
\end{figure}

To accurately measure the line wavenumbers in the FFP spectrum, a scan of the reference C$_2$H$_2$ cell was taken directly before and after each FFP scan to precisely determine the wavenumber correction factor of the spectrograph. This was done without replacing the output fiber of the SLED or the illumination fiber to the FTS. A high S/N scan of the FFP using the wavelength solution from the C$_2$H$_2$ cell is shown in Figure~\ref{fig:nist_scan}. This scan allowed for precise determination of the FFP line spacing across the SLED continuum wavelengths. The line spacing varies in regions where the interferometric line intensity deviates significantly from the SLED continuum curve (see Figure~\ref{fig:line_spacing_err}). The origins of this deviation are unclear at this point, though the likely cause is non-uniform phase dispersion of the FFP near these wavelengths. This is not expected to effect the device stability, so long as the dispersion remains static over time.

A known issue with these SLEDs is periodic spectral ringing, possibly due to reflection fringing at internal fiber mating interfaces. This ripple effect is imprinted on the FFP transmission function (see Figure~\ref{fig:ffp_sled}), though this has little effect on the FFP line profiles.
Repeated tests using the FFP and NIST FTS are planned for the second half of 2012. These tests are aimed at estimating day-to-day line stability to high accuracy, characterizing phase dispersion effects on line centers and intensities, and determining line profiles for the entire operating range of the FFP (1260 - 1640 nm). Upcoming tests will utilize a 1 W super-continuum source with a more stabilized pulsation rate. This should greatly reduce noise associated with varying repetition rate, and allow for a complete scan of the FFP transmission lines from 900 - 2000 nm.

\section{Conclusions}
\begin{table}
\begin{center}
\caption{Summary of components used in testbed fiber Fabry-Perot system.}

\begin{tabular}{cc}
\hline
\hline
Component	&	Cost	\\
\hline
SuperK Compact\footnotemark[1]	&	\$25k	\\
Micron Optics\footnotemark[1] FFP	&	$\sim\$$10k	\\
Stanford Research Systems\footnotemark[1] TEC contoller	&	\$5k	\\
Testing equipment (power meter, NIR diodes, safety glasses, precision resistor)		&	\$5k	\\
Acronym\footnotemark[1] 1550 narrowband fiber laser	&	\$1k \\
\hline
\label{tab:ffp_components}
\end{tabular}

\end{center}

\end{table}
This device represents a simple, economic solution to an important issue in precise Doppler spectroscopy. Combining astronomical instrumentation with commercial photonic devices can yield a wealth of scientific return. The output of the FFP is similar to that of an LFC but at a small fraction of the cost. Few NIR wavelength references exist that are precise to the m s$^{-1}$ level, insensitive to both vibration and pressure changes, rigid in design, and relatively inexpensive to construct. The FFP satisfies all these goals in a compact unit. Other high resolution NIR fiber spectrographs could easily be converted into precise radial velocimeters with the addition of such a device.
A full list of all components used in testing and commissioning the device is shown in Table~\ref{tab:ffp_components}. Our goal here is to present the initial performance report and discuss the properties of the FFP device, as well as the unique illumination source used. Further testing will be required to fully characterize both the stability, and intrinsic transmission spectrum of the FFP.
A detailed analysis of the APOGEE data, as well as new results from the upcoming NIST tests will be presented in a separate refereed paper.

\acknowledgments{This work was partially supported by the Center for Exoplanets and Habitable Worlds, which is supported by the Pennsylvania State University, the Eberly College of Science, and the Pennsylvania Space Grant Consortium. We acknowledge support from NSF grant AST-1006676, the NASA Astrobiology Institute (NAI), and PSARC. This research was performed while SLR held a National Research Council Research Associateship Award at NIST. NSO/Kitt Peak FTS data used here were produced by NSF/NOAO.

Funding for SDSS-III has been provided by the Alfred P. Sloan Foundation, the Participating Institutions, the National Science Foundation, and the U.S. Department of Energy Office of Science. The SDSS-III web site is {http://www.sdss3.org/}.

SDSS-III is managed by the Astrophysical Research Consortium for the Participating Institutions of the SDSS-III Collaboration including the University of Arizona, the Brazilian Participation Group, Brookhaven National Laboratory, University of Cambridge, Carnegie Mellon University, University of Florida, the French Participation Group, the German Participation Group, Harvard University, the Instituto de Astrofisica de Canarias, the Michigan State/Notre Dame/JINA Participation Group, Johns Hopkins University, Lawrence Berkeley National Laboratory, Max Planck Institute for Astrophysics, Max Planck Institute for Extraterrestrial Physics, New Mexico State University, New York University, Ohio State University, Pennsylvania State University, University of Portsmouth, Princeton University, the Spanish Participation Group, University of Tokyo, University of Utah, Vanderbilt University, University of Virginia, University of Washington, and Yale University.

\bibliography{FFP_SPIE}   %>>>> bibliography data in report.bib
\bibliographystyle{spiebib}   %>>>> makes bibtex use spiebib.bst

\end{document}